\documentclass{article}

\usepackage[dvips]{epsfig}
\usepackage{rotating}

\begin{document}
 
\title{
The Study of Correlation Structures of DNA Sequences:  \\
  A Critical Review
\vspace{1in}
\author{Wentian Li\thanks{email: wli@linkage.rockefeller.edu} \\
{\small \sl Laboratory of Statistical Genetics, Rockefeller University,}\\
{\small \sl 1230 York Avenue, Box 192, New York, NY 10021}\\ 
} 
\date{to be published in the special issue of Computer \& Chemistry\\
(March, 1997)}
} 
\maketitle    
\markboth{\sl W. Li}{\sl W. Li}

\begin{abstract}
The study of correlation structure in the primary sequences of DNA
is reviewed. The issues reviewed include: symmetries among 16
base-base correlation functions, accurate estimation of correlation
measures, the relationship between $1/f$ and Lorentzian spectra,
heterogeneity in DNA sequences, different modeling strategies of the 
correlation structure of DNA sequences, the difference of correlation 
structure between coding and non-coding regions (besides the period-3 
pattern), and source of broad distribution of domain sizes. Although
some of the results remain controversial, a body of work on this 
topic constitutes a good starting point for future studies.
\end{abstract}

\newpage


\section*{0. Introduction}

There is a long-standing interest in knowing and understanding the 
correlation between bases in DNA sequences. Before the human genome 
project era, there were no long, continuous DNA sequences. The study 
then concerned either the nearest-neighbor base-base correlations
[Josse, et. al., 1961] or the base density heterogeneity
in digested DNA segments [Sueoka, 1959]. A more complete characterization
of correlation between base pairs at both short and long distances
became possible only as long DNA sequences became more commonly available.

Not all studies of a complete characterization of correlation
structure of DNA sequences were motivated by biology. Rather, many
such studies were motivated by the issues of mathematical modeling, 
dynamical systems, stochastic processes, and noise.  Perhaps due to 
this reason,  this study has not yet become part of the toolbox in 
the ``mainstream" DNA sequence analysis.

This review is an attempt to summarize the current status of this
study. There are at least two goals for this review. First, there have 
been disagreements on the result of correlation structure in DNA sequences. 
Due to this uncertainty of what the actual result is, some people still 
believe that DNA sequences do not exhibit any feature which cannot be 
explained by the basic stochastic processes such as random sequence or Markov 
chain - with the first process having no correlation and the second one 
having only short-range correlations. Resolving this disagreement can
be straightforward once everybody agrees to use the same measure of
correlation, use the same estimator, and apply this estimator of the 
correlation to the same sequence.

The second goal is to promote more biologically-motivated study of
correlation structure of DNA sequences. Although this paper does not
accomplish this task, the intention is to at least raise the issue. Most
of the current studies of correlation in DNA sequences are based on base-base
statistical correlations. This base-base correlation may not be a powerful 
way to reveal the correlation on a global scale or between larger units.
Using an analogy of the natural language texts: statistical correlation 
between letters in an English text rarely reveals correlations at 
the syntax level in a sentence, or the correlation between
sections, or the overall organization of the text, because the linguistically
meaningful units are words, sentences, paragraphs, instead of letters.

Besides the above note that most correlations studied are base-base
correlation, another caution is that these are {\sl statistical} 
correlations. A statistical correlation between two events exists if
and only if the joint probability of the two events is not equal
to the product of the two probabilities for each event.  In other analysis
of correlations, such as the one between a risk factor (smoking) and a
disease (cancer), the distinction is sometimes made between 
{\sl causal correlation} and {\sl spurious correlation}. The causal 
correlation is the correlation between true cause and the effect.
The spurious correlation is the non-causal part of the statistical
correlation. Without a definition of the causal base-base correlation, 
it is not clear whether the above distinction is applicable here.

The claim of (base-base, statistical) correlation at long distances
in DNA sequences is still a few steps away from finding an organization
principle of the genome.  The quote from [Ohno,1993] for example, 
 ``$\dots$ this would suggest the existence of a `grand design' 
in the construction of DNA sequences $\dots$", 
represents such a misunderstanding. In order to discover a true
``organization principle", one might start from the biologically
meaningful units, arrange both the structurally functional regions
and the genetically functional regions in place, and see how the
organization helps the living of the cell.

There is another confusion related to this study. The term 
``long-range" in the ``long-range (base-base, statistical)
correlation" was meant to be longer than (1) 3-6 bases [Li, 1992];
or (2) 800 bases [Li and Kaneko, 1992a]; or (3) 1 -10 kb
(1 kb = 1000 bases) [Peng, et. al. 1992]. A recent analysis of complete
DNA sequences of budding yeast chromosomes ({\sl Saccharomyces cerevisiae}) 
also suggested that the correlation exists up to more than 10 kb, but is 
absent at 100 kb range (W.Li, unpublished results).
On the other hand, the ``long-range" in  ``long-range physical mapping"
typically means the range covered by the current
physical mapping techniques  is as large as 
100 kb - 1 Mb (1 Mb = 1,000,000 bases). Clearly, what is 
considered to be long is relative to what is considered to 
be short. For this reason, the term ``long-range" will be used with
caution in this review. More often, the term ``correlation structure" 
is used instead to indicate the pattern of correlation at all distances.

This paper is organized as follows: section 1 describes the original
motivation for a complete characterization of correlation structure 
in DNA sequences; section 2 reviews both the measures used in
characterizing correlation structure (in spatial as well as spectral 
domain) and some published results on these correlation structures
of DNA sequences; two special subsections are included: one on the
estimation of correlations (sec. 2.2), and another on characterizing
heterogeneity of DNA sequences (sec. 2.7); section 3 discusses different
approaches to the modeling of the correlation structure of DNA sequences; 
and finally, section 4 discusses some biological issues related to 
the correlation study.  

A review of this sort is inevitably biased, in the sense that the 
material chosen reflects the author's interests and experience, and
the opinions expressed are mainly the author's. Nevertheless, I intend to 
make the review as balanced and as fair as possible. An early review 
covers the study on this topic before 1993 can be found in 
[Li, et. al., 1994].

\section{Background}

As early as the sixties, there were attempts to characterize the 
statistical properties of DNA sequences: for example,  the correlations 
between the nearest-neighbor bases 
[Josse, et. al., 1961; Gatlin,1966] and the heterogeneity 
of base density in fragmented DNAs [Sueoka, 1959]. Statistical 
regularities were used to detect coding regions 
[Shulman, et. al., 1981; Shepherd, 1981a,1981b; Staden and McLachlan, 
1982; Fickett, 1982], and to study the nucleosome formation 
[Trifonov and Sussman, 1980].
All these studies focus on a particular aspect of the correlation
structure of DNA sequences in relation to a particular biological question.

For mathematically-oriented researchers, a DNA sequence might be considered 
as a string of symbols whose correlation structure can be characterized 
completely by all possible base-base correlation functions or their 
corresponding power spectra. For people interested in dynamical systems, 
there is also a ``dynamical" aspect to this otherwise static problem: 
the change of DNA sequences (even without a reference to the natural 
selection) can be considered as the updating of the symbolic sequences. 
This updating, once described by a mathematical model, 
can be studied easily on a computer. One question is which model will 
generate what types of correlation structure in a symbolic sequence.

A few years ago, I was studying such a problem. The dynamical system
being studied which updates a symbolic sequence is the so-called 
``cellular automata" [Toffoli and Margolus, 1987; Wolfram,1986]. 
These systems update sequence locally. The question asked 
was whether these locally-operated systems are able to generate 
global correlations. The answer turned out to be negative for 
simple cellular automata [Li, 1987]. 

Two events turned my attention to DNA sequences. One was a publication
trying to model the DNA evolution by cellular automata
[Burks and Farmer, 1984]. Another was the discovery that long-range
correlation in a sequence could be easily generated if the sequence
is allowed to increase [Li, 1989]. In what I called the
``expansion-modification system" [Li, 1989; 1991], there are only two 
processes: ``expansion" and ``modification". When the 
expansion rate far exceeds the modification rate, the sequences generated
by the system exhibit a long-range correlation called ``1/f spectrum"
(to be discussed more in the sec. 2.4). It was pointed out to me
(K. Kaneko, private communication, 1988-89) that the expansion-modification
model I was studying was reminiscent of the biological processes: 
expansion being the base or oligonucleotide duplication process,
and modification being the point mutation. This led us to a study 
of the correlation structure
in DNA sequences [Li, 1992; Li and Kaneko, 1992a].
 
Other groups interested in studying correlation structure in DNA sequences
were probably motivated by different reasons. For example, in 
[Peng, et. al. 1992], the interest was perhaps to compare 
power-law functions observed in a representation of DNA sequences to 
other self-similar phenomena in nature. The interest in [Voss, 1992] 
was to compare the 1/f spectra observed in DNA sequences to other 
1/f noise in nature such as musical signals [Voss and Clarke, 1975]. 
[Voss, 1992] also suggested that 
the information storage in DNA sequences is between efficient (as in 
random sequences) and redundant (as in repetitive sequences).

\section{Correlation structure of DNA sequences}

\subsection{Direct measure of correlation}

The basic measure of correlation among bases in a DNA sequence
are the 16 correlation functions between all 16 possible base-pairs:
\begin{equation}
\{ \Gamma_{\alpha \beta}(d) \}\equiv
\{ \Gamma_{AA}(d), \Gamma_{AC}(d), \dots, \Gamma_{GT}(d), \Gamma_{TT}(d) \}
\end{equation}
each defined as the correlation between nucleotide $\alpha$ 
and another nucleotide $\beta$ separated by a distance $d$:
\begin{equation}
\Gamma_{\alpha \beta}(d) \equiv P_{\alpha \beta}(d) - 
P_{\alpha\cdot} P_{\cdot \beta}
\hspace{0.4in}
\mbox{     $\alpha,\beta = \{ A, C,G,T \}$},
\end{equation}
where $P_{\alpha\beta}(d)$ is the joint probability of observing
$\alpha$ and $\beta$ separated by a distance $d$,
$P_{\alpha \cdot} \equiv \sum_{\beta'} P_{\alpha\beta'}(d)$
and $P_{\cdot \beta} \equiv \sum_{\alpha'} P_{\alpha'\beta}(d)$
are the density for nucleotide $\alpha$ and $\beta$, respectively.
Because of the relations $\sum_{\alpha} \Gamma_{\alpha\beta}(d) 
= \sum_{\beta} \Gamma_{\alpha\beta}(d) =0$, the number of independent
correlation functions (not yet considering any other symmetries) 
is actually 9 [Herzel and Gro\ss e, 1995].

Following are comments concerning these correlation functions:
\begin{enumerate}
\item
{\bf Statistics are taken along a sequence}:
Suppose ${\bf 1}_\alpha(i)$ is 1 when the nucleotide at
position $i$ is $\alpha$, and 0 otherwise. 
$N_{\alpha \beta}(d) \equiv \sum_{i=1}^{N-d} 
{\bf 1}_\alpha(i) {\bf 1}_\beta(i+d)$
is the count of $(\alpha,\beta)$ pair separated by a distance $d$
(the sampling is stopped when the second base reaches the end of
the sequence).  If one uses the periodic boundary condition, 
$N_{\alpha \beta}(d)^c \equiv \sum_{i=1}^{N}
{\bf 1}_\alpha(i) {\bf 1}_\beta(i+d)$ (where $c$ indicating ``circular").

$P_{\alpha \beta}(d)$ is estimated {\sl along} the sequence, either
by $\widehat{P_{\alpha \beta}(d)}=N_{\alpha \beta}(d)/(N-d)$ or by
$\widehat{P_{\alpha \beta}(d)^c}=N_{\alpha \beta}(d)^c/N$.
(Note: since the number of counts can either be $N$ or $N-d$,
depending on whether the periodic boundary condition is used or
not, the symbol $N$ used in Eqs.(\ref{eq4},\ref{eq5},\ref{eq7},\ref{eq8},
\ref{eq9}, \ref{eq13}, \ref{eq15}) can be either $N$ or $N-d$.)
Because statistics are taken along the sequence, all heterogeneity in 
$P_{\alpha \beta}(d)$ along the sequence will be averaged out.
For example, $N_{\alpha\beta}(d)$ counts may mainly come from
one region of the sequence,  or from throughout the sequence.
But from the number $N_{\alpha\beta}(d)$, we do not know which is the case.
Consequently, $\{ \Gamma_{\alpha \beta}(d) \}$ at a fixed distance $d$ does 
not contain information on heterogeneity. Nevertheless,
{\sl $\{ \Gamma_{\alpha \beta}(d) \}$ as a function of $d$ does reveal
heterogeneity in the sequence}.

\item
{\bf Strand complementarity}:
DNA sequences are double-stranded with nucleotides on one strand complementary
to those on the other. As a result, $\Gamma_{\alpha \beta}(d)$ on one strand
(in 5' $\rightarrow$ 3' direction) is exactly the same with the
$\Gamma_{\bar{\beta}\bar{\alpha}}(d)^{opposite}$ on the opposite strand 
in the opposite direction (but also in 5' $\rightarrow$ 3' direction
viewed from that strand). For example, in Fig.1,
$\Gamma_{CT} = \Gamma_{AG}^{opposite}$

\item
{\bf Strand symmetry}:
It was observed that $\Gamma_{\alpha \beta}(d)$ on one strand is 
{\sl approximately} equal to $\Gamma_{\alpha \beta}(d)^{opposite}$ 
on the opposite strand
in the opposite direction. This idea of ``strand symmetry" was 
suggested in [Fickett, et. al., 1992], though they actually suggested
the symmetry for base density and density-density correlation,
not base-base correlation as explicitly written here. Combining
strand symmetry with strand complementarity, we have
$\Gamma_{\alpha \beta}(d) \approx \Gamma_{\bar{\beta}\bar{\alpha}}(d)$ 
on one strand. For example, in Fig.1, 
$\Gamma_{CT} \approx \Gamma_{AG}$. This approximate symmetry reduces
the number of independent correlation functions from 9 to 6.

There were suggestions of other symmetries. For example, in
[Teitelman and  Eeckman, 1996], it was suggested that correlation
is almost the same under simultaneous
$A \rightarrow T$ and $T \rightarrow A$ transformation 
(e.g. $\Gamma_{AG} \approx \Gamma_{TG}$). This symmetry reduces the
number of independent correlation functions from 9 to 5. If the
correlation matrix is invariant under simultaneous
$A \rightarrow T$, $T \rightarrow A$, 
$C \rightarrow G$, $G \rightarrow C$ transformation, the
number of independent correlations is reduced from 9 to 3.

Fig. 2 shows the 16 correlation functions $\Gamma_{\alpha\beta}(d)$
for $d$ from 1 to 1000, determined from the budding yeast 
chromosome 1 (using the Bayesian estimator as will 
be discussed in the next section).  It is clearly seen that 
$\Gamma_{AA}(d) \approx \Gamma_{TT}(d)$ and $\Gamma_{CC}(d) \approx 
\Gamma_{GG}(d)$, with all other cross-correlations
roughly similar to each other. 

\item
{\bf Relative contribution to the correlation from different pairs}:
Among the 9 independent correlation functions, some contribute
more to the overall correlation than others. 
In [Teitelman and  Eeckman, 1996], the data showed that
$\Gamma_{AA}(d) (\approx \Gamma_{TT}(d))$ is the largest among all
correlation functions. $\Gamma_{GG}(d) (\approx \Gamma_{CC}(d))$
is the second largest.  The calculation from budding yeast 
chromosome 1 (Fig.2) confirms this.

Also note that the correlations between the same nucleotide (e.g., 
$\Gamma_{AA}$, $\Gamma_{GG}$) are always positive, and those between 
different nucleotides (e.g. $\Gamma_{AT}$, $\Gamma_{GC}$ ) are 
usually negative (M. Zhang, unpublished draft, 1992; also see Fig.2).
This can actually be explained in the case of simple domains structure
(appendix A of [Li, et. al., 1994]).

\item
{\bf Average over all correlation functions}:
Rather than calculate all independent correlation functions, 
we can define one measure which takes into account all individual
correlation functions. For example, the mutual information function
is defined as ([Shannon, 1948; Li, 1990])
(the base of the log term can also be 4, 10, or $e$, rather than 2):
\begin{equation}
M(d) \equiv \sum_{\alpha \beta} P_{\alpha \beta}(d) \log_2 
\frac{P_{\alpha \beta}(d)}{ P_{\alpha \cdot} P_{\cdot \beta} }
=\sum_{\alpha \beta} P_{\alpha \beta}(d) 
(\log_2 P_{\alpha \beta}(d) - \log_2 (P_{\alpha \cdot} P_{\cdot \beta} ))
\end{equation}
or $X^2$ (see, e.g., sec.13.4 of [Rice, 1995]):
\begin{equation}
\label{eq4}
X^2(d) = N \sum_{\alpha \beta} 
\frac{ (P_{\alpha \beta}(d) - P_{\alpha \cdot} P_{\cdot \beta})^2}
{P_{\alpha \cdot} P_{\cdot \beta}}.
\end{equation}
$X^2$ should obey the $\chi^2$ distribution with 9 degrees of freedom
under the zero-correlation hypothesis [Kullback, 1959].

The sufficient and necessary condition for no correlation between
two bases is that all correlation functions are equal to 0.
It is equivalent to $M(d)=0$ or $X^2(d)=0$.  In fact, in a first
approximation,  $M(d) \approx (1/2N) X^2(d)$ [Li, 1990;
Herzel and Gro\ss e, 1995].
\end{enumerate}

\subsection{Estimation of correlations from a sequence with finite length}

Since the correlation at longer distances is typically small, it is
important to use the best possible estimator to measure the correlation. 
Otherwise, the error due to a finite sample size can be as large as 
the correlation value itself. We discuss three estimators here:

\begin{enumerate}

\item
{\bf Frequency-count estimator}:
It is usually assumed that the best way to estimate
the  probability of an event
$a$ ($P_{a}$) is to divide the number of count for $a$ ($N_{a}$) by the total 
number of count ($N$), i.e., the frequency-count  estimator:
\begin{equation}
\label{eq5}
(\widehat{P_a})_{freq} = \frac{N_a}{N}.
\end{equation} 
This estimator is obtained
by maximizing (here ``data" is $N_a$ and ``parameter" is $P_a$)
\begin{equation}
Likelihood \equiv Prob(data|parameter) 
\end{equation}
to get the parameter. In our case, when
$Prob(N_a|P_a) \sim P_a^{N_a} (1-P_a)^{N-N_a}$
is maximized with respect to $P_a$, we obtain Eq.(\ref{eq5}).

Although $(\widehat{P_a})_{freq}$ is an unbiased estimator of
$P_a$, inserting these estimators to a derived quantity of $P_a$ may
not be an unbiased estimator of that quantity. For example,
if entropy $H = -\sum_\alpha P_\alpha \log_2(P_\alpha) $ is estimated
by $\widehat{H} = -\sum_\alpha N_\alpha/N \log_2(N_\alpha/N)$, there
is an underestimation [Basharin, 1959; 
Herzel, 1988; Herzel and Gro\ss e, 1995; Gro\ss e, 1995]. Similarly, 
\begin{equation}
\label{eq7}
\widehat{M(d)}_{freq} = \sum_{\alpha\beta} \frac{N_{\alpha\beta}(d)}{N}
\log_2 \frac{N_{\alpha\beta}(d)N}
{N_{\alpha\cdot}  N_{\cdot\beta}}
\end{equation}
usually overestimates $M(d)$ [Li, 1990; Herzel and Gro\ss e, 1995; 
Gro\ss e, 1995]. (Note: the eq.(5.10) in [Li, 1990] should be 
$(K-1)^2/2N$ instead of $K(K-2)/2N$.) The bias in the frequency-count 
estimator can be corrected when it is approximately calculated. 
Nevertheless, the variance of the estimator around the true 
value is not reduced by this correction.

The frequency-count estimator for correlation 
function $\Gamma_{\alpha\beta}(d)$ would be:
\begin{equation}
\label{eq8}
\widehat{\Gamma_{\alpha\beta}(d)}_{freq}
= \frac{N_{\alpha\beta}(d)}{N} - \frac{ N_{\alpha \cdot}}{N} 
\cdot \frac{ N_{\cdot \beta}}{N}
\end{equation}

\item
{\bf Indirect Bayesian estimator}:
A completely different type of estimator is the 
Bayesian's estimator (see, e.g. sec.15.2.4 [Rice,1995]). The Bayesian
estimator of an event out of $K$ possible states is:
\begin{equation}
\label{eq9}
(\widehat{P_a})_{bayesian} = \frac{N_a+1}{N+K}.
\end{equation}
This estimator is obtained by first calculating the posterior
probability for the parameter:
\begin{equation}
Posterior \equiv  Prob(parameter|data),
\end{equation}
then using this probability as the weight to calculate the average 
value of the parameter. In the case of two possible outcomes, 
\begin{equation}
Prob(P_a|N_a)= \frac{Prob(N_a|P_a) Prob(P_a)}{ Prob(N_a)} =
\frac{ P_a^{N_a} (1-P_a)^{N-N_a} Prob(P_a)}{
 \int_{P_a=0}^1 P_a^{N_a} (1-P_a)^{N-N_a} Prob(P_a) dP_a},
\end{equation}
and assuming $Prob(P_a) =const.$ ( i.e., uniform prior distribution),
we have 
\begin{equation}
(\widehat{P_a})_{bayesian} 
= \int_{P_a =0}^1 Prob(P_a|N_a) P_a d P_a
= \frac{\int_{P_a=0}^1 P_a^{N_a+1} (1-P_a)^{N-N_a} dP_a}
{ \int_{P_a=0}^1 P_a^{N_a} (1-P_a)^{N-N_a} dP_a}, 
\end{equation}
which leads to Eq.(\ref{eq9}) (set $K=2$).

So the second estimator of a derived quantity of the probabilities
is obtained by inserting the Bayesian estimator of the probabilities to the
function [P. Grassberger, unpublished result, 1994]. This ``indirect"
Bayesian estimator reduces the variance, but there is still a substantial
bias [Gro\ss e, 1996]. The indirect Bayesian estimator for
correlation function $\Gamma_{\alpha\beta}(d)$ is (for $K=4$):
\begin{equation}
\label{eq13}
(\widehat{\Gamma_{\alpha\beta}(d)})_{ind\_bayesien}
= \frac{N_{\alpha\beta}(d)+1}{N+16} - 
\frac{ N_{\alpha \cdot}+4}{N+16} 
\cdot \frac{N_{\cdot \beta}+4}{N+16}
\end{equation}

\item
{\bf Direct Bayesian estimator}:
A direct Bayesian estimator for a quantity does not rely on the
Bayesian estimator of the probabilities, but a direct average
of the quantity using the posterior probability. For example,
the direct Bayesian estimator for correlation function 
$\Gamma_{\alpha\beta}(d)$ is:
\begin{equation}
(\widehat{\Gamma_{\alpha\beta}(d)})_{bayesian}
= \frac{\int_{\{ P_{\alpha'\beta'} \}}
\left\{  P_{\alpha\beta} -P_{\alpha \cdot} P_{\cdot \beta} \right\}
\prod_{\alpha'\beta'}
P_{\alpha'\beta'}^{N_{\alpha'\beta'}} Prob(P_{\alpha'\beta'}) 
d \{ P_{\alpha'\beta'} \} }
{\int_{ \{ P_{\alpha'\beta'} \}}
\prod_{\alpha'\beta'} P_{\alpha'\beta'}^{N_{\alpha'\beta'}} 
Prob(P_{\alpha'\beta'}) d \{ P_{\alpha'\beta'} \} }
\end{equation}
For $K=4$, it can be calculated to be 
[Wolpert and Wolf, 1995; Gro\ss e, 1995] (using
the multinomial distribution, and notice $\sum_{\alpha\beta} P_{\alpha\beta}=1$
and $\sum_{\alpha\beta} N_{\alpha\beta} =N$ ):
\begin{equation}
\label{eq15}
(\widehat{\Gamma_{\alpha\beta}(d)})_{bayesian}
= \frac{N_{\alpha\beta}(d) +1}{ N+17}
-\frac{N_{\alpha \cdot} +4}{N+16} \cdot \frac{N_{\cdot \beta} +4}{N+17}
\end{equation}
Although the direct Bayesian estimator is still biased, the bias 
is much smaller than the frequency-count  estimator and the indirect 
Bayesian estimator, and the variance is also reduced 
as compared with the frequency-count  estimator [Gro\ss e, 1996].

\end{enumerate}

\subsection{Direct measure of correlation in DNA sequences}

It is straightforward to apply the direct measure of correlation to DNA 
sequences, as was done, for example, in [Shepherd, 1981a, 1981b; 
Fickett,1982; Konopka and Smythers, 1987; Konopka, et. al. 1987;
Arqu\`{e}s and Michel, 1987;  Li, 1992; Li and Kaneko, 1992a; 
Mani, 1992]. Besides the
well-known period-3 oscillation in coding regions,  a calculation of 
the complete correlation function gives us more information, such as
whether the correlation function decays as a power-law function
or as an exponential function or in-between. Here are some comments:
\begin{enumerate}
\item
{\bf Different sequences may exhibit different correlation functions}:
When a correlation function is calculated for an individual DNA sequence,
its form may be different from one sequence to another. In [Li, 1992],
for example, it is shown clearly that the $M(d)$'s for the 5 human exon 
sequences are different from the 5 human intron sequences. In another
example, the bacteriophage lambda sequence used in [Karlin and Brendel, 1993] 
has $1/f^2$ spectrum [Li, et.al., 1994]
(to be discussed in sec. 2.4-2.5)  whereas the budding yeast
chromosome 3 sequence exhibits 1/f-like spectrum [Li, et. al., 1994].
As a result, any conclusion from the analysis of one sequence should
be taken with care when generalized to another sequence.

\item
{\bf Correlation function obtained from the whole sequence may be
different from that obtained from a sub-sequence}:
Closely related to the comment \#1 in sec. 2.1, since the statistic
is sampled along the sequence, a correlation function based on the
statistic sampled from the whole sequence may differ from that based
on a statistic sampled from a sub-sequence. There have been 
statements such as: ``no long-range correlations are found in any of 
the studied DNA sequences" [Azbel, 1995], while the study was only
carried out at the subsequence level. 

It is also possible that a correlation is present at the subsequence
but will not be extended to longer distances when the whole sequence
is analyzed. In fact, no one currently has the sequence of a complete human
chromosome thus the correlation structure at the length scale of
the whole human chromosome is unknown.

\item
{\bf Correlation function from one sequence may be different from
that averaged over many sequences}:

This is yet another seemingly trivial statement but can be overlooked
in a debate on the nature of long-range correlation in DNA sequences.
When the correlation functions from many sequences are averaged, the 
one with the slowest decay rate dominates at the long distances. 
If the correlation function
in each sequence decays exponentially but with different rates (i.e.,
different correlation lengths), the averaged correlation function may
decay as an non-exponential function, such as a power-law function
(it is closely related to the comment \#6 in the next section).
\end{enumerate}

\subsection{Spectral analysis}

Power spectra via Fourier transform
(see, e.g., [Percival and Walden, 1993]) is widely used in time series
analysis. The estimator of the power spectrum for $(\alpha, \beta)$ pair 
is defined as
\begin{equation} 
\widehat{S_{\alpha \beta}(k)}
\equiv \left(\frac{1}{N} \sum_{j=1}^{N} {\bf 1}_\alpha(j) 
e^{2 \pi i j (k/N) }\right)
\left( \frac{1}{N} \sum_{j'=1}^{N} {\bf 1}_\beta(j') 
e^{2 \pi i j' (k/N) }\right)^*
\end{equation} 
where ${\bf 1}_\alpha(j)$ is 1 if the symbol at position $j$ 
is $\alpha$ and 0 otherwise;  the $*$ is the complex conjugate.
The frequency $f$ is defined as $f=2 \pi k/N$. Although
$k$ can range from 0 to $N-1$, due to the mirror symmetry around
$k=N/2$, typically only the spectrum from $k=0$ to $k=N/2$ is displayed.

There is a one-to-one correspondence between 
the power spectrum $\widehat{S_{\alpha \beta}(k)}$ and the circular
correlation function $\widehat{\Gamma_{\alpha \beta}(d)}^c$ 
[Chechetkin and Turygin, 1994] (Note: strictly speaking,
no such relation holds between $\widehat{S_{\alpha \beta}(k)}$ and
the ``non-circular" $\Gamma_{\alpha \beta}(d)$):
\begin{equation}
\label{eq17}
\widehat{S_{\alpha \beta}(k)} = \frac{1}{N} \sum_{d=0}^{N-1} 
\widehat{\Gamma_{\alpha \beta}(d)}^{c}
e^{ - i 2 \pi d (k/N) }
= \frac{\widehat{\Gamma_{\alpha\beta}(0)}^c}{N}
+ \frac{2}{N} \sum_{d=1}^{N/2} \widehat{\Gamma_{\alpha \beta}(d)}^c 
\cos( 2 \pi d k/N)
\end{equation}
The second expression is due to the mirror symmetry around $k =N/2$.

Since power spectrum and correlation function are two representations
of the same correlation structure, power spectrum does not provide
any new information which is not described by the correlation function.
Nevertheless, the visual representation of a power spectrum sometimes
can more easily reveal patterns which are harder to discern in 
the correlation function.  The following comments relate to
power spectra:
\begin{enumerate}
\item
{\bf Averaged power spectrum}:
Similar to the case of correlation functions, there are many
ways to average or sum power spectra $S_{\alpha\beta}$. For example,
\begin{equation}
S_{ave1}(k) \equiv \sum_{\alpha \beta} |S_{\alpha\beta}(k)|
\end{equation}
or (e.g., [Voss, 1992; Li, et. al., 1994])
\begin{equation}
S_{ave2}(k) \equiv \sum_{\alpha} S_{\alpha\alpha}(k)
\end{equation}
or to assign four nucleotides to the four vertices of a tetrahedron,
use the 3 coordinates of a vertex to represent a nucleotide, then
calculate the sum of power spectra of sequences from each coordinate
[Silverman and Linsker, 1986; Li and Kaneko, 1992a]. It seems that
different projections of the power spectrum do not alter the 
general shape of the power spectrum.

\item
{\bf Exponential-decaying correlation functions correspond to
Lorentzian spectrum with a $1/f^2$ tail}:
In Eq.(\ref{eq17}), if the (circular) correlation function decays
exponentially: $\Gamma(d)^c \sim \lambda^d$,  where $  0 < \lambda < 1$,
it can be shown that in the limit of $N \rightarrow \infty$
[Bor\u{s}tnik, et. al.,  1994] (if $N$ is finite,
there are many more terms in the expression, but the main conclusion
remains the same):
\begin{equation}
S(k) \sim  const. + \frac{ \cos (2 \pi k/N) - \lambda}
{(1-\lambda)^2/2\lambda +(1 - \cos( 2 \pi k/N))}
\end{equation}
Using the Taylor expansion of $\cos(x) \approx 1 - x^2/2$, the above
expression can be approximated as
\begin{equation}
S(f) \sim \frac{const.}{ const. + f^2}
\end{equation}
where $f = 2 \pi k/N$ is the frequency. This spectral form is
called a Lorentzian spectrum. 

If the correlation function does not decay monotonically, but
is oscillational, the Lorentzian spectrum will be centered around
a non-zero frequency due to the periodicity.

\item
{\bf Step functions exhibit $1/f^2$ power spectrum}:
It is very easy to show that the correlation function of a step function
(e.g., $x(i)=1$ if $ 1 \le i \le N/2$ and $x(i)=0$ if $N/2 < i \le N$)
decays linearly (see, e.g., appendix A of [Li, 1991] and appendix A
of [Li, et. al., 1994]). By using  Eq.(\ref{eq17}), the
corresponding power spectrum contains a $1/f^2$ term. The implication
of this almost trivial result is that for DNA sequences which
are C+G-rich on one half but C+G-poor on the other half, the
power spectrum is expected to be of the form $S(f) \sim 1/f^2$. 

\item
{\bf When both the correlation function and  the power spectrum
are power-law functions}:
Since the Fourier transform of a power-law function is still
a power-law function, we have (using Eq.(\ref{eq17}) in the
$N \rightarrow \infty$ limit and approximate the sum by an
integral)
\begin{equation}
(S(f) \sim )\frac{1}{f^b} \sim  \int_{x=1}^{\infty} \frac{1}{x^a} \cos(xf) dx
(\sim \int_{x=1}^\infty \Gamma(x) \cos(xf) dx )
\end{equation}
Suppose we change $f$ to $kf$
\begin{equation}
\frac{1}{(kf)^b} \sim  \int_1^{\infty} \frac{1}{x^a} \cos(xkf) dx
=  \frac{k^a}{k}\int_1^{\infty} \frac{1}{ (kx)^a} \cos( (kx) f) d (kx)
\sim \frac{1}{k^{1-a}} \frac{1}{f^b}
\end{equation}
In other words, $b \approx 1-a$. This ``scaling argument" or
``dimension analysis" has been frequently used in physics. The step
function discussed in comment \#3 confirms this relationship 
since $a=-1$ and $b=2$.

Caution should be taken for many real situations. For example,
there can be cutoffs of the power-law function at both high- and
low- frequencies [Theiler, 1991].  It is also possible that 
the power spectrum is only a stepwise power-law function.

\item
{\bf 1/f spectra (1/f noise, 1/f fluctuation, flicker noise)}:
A particular interesting situation is when $b \approx 1$ which implies
$a \approx 0$, or the correlation function decay to zero very slowly.
What makes ``$1/f$ spectra" or ``$1/f$ noise" interesting is
that this type of fluctuation is very common in nature [Press, 1978].
1/f noise was perhaps first observed and studied in the noise of electronic 
systems [Johnson, 1925; Schottky, 1926], but it appears in numerous 
other phenomena, ranging from star luminosity to traffic flow (an online 
bibiliography on 1/f noise can be found at URL: 
{\sl http://linkage.rockefeller.edu/wli/1fnoise}).

\item
{\bf 1/f spectra as a superposition of Lorentzian spectra}:
A natural and popular explanation of 1/f noise is that
these are superpositions of many series with exponentially-decaying
correlation function, each is sampled from a broad distribution of the
correlation length $\tau$ (in $\Gamma(x) \sim e^{-x/\tau}$). 
One specific probability density function of the correlation length $\tau$ 
that leads to 1/f spectra is
\begin{equation}
\label{eq24}
g(\tau) d\tau =
\left\{
\begin{array}{ll}
 \frac{1}{\ln(\tau_{max}/\tau_{min})} \frac{1}{\tau}  d \tau &
\mbox{ if $ \tau_{min} < \tau < \tau_{max}$ } \\
0 & \mbox{ otherwise}
\end{array}
\right.
\end{equation}
[van der Ziel, 1950].  The log-normal distribution can 
approximately lead to the $1/\tau$ distribution
[Montroll and Shlesinger, 1982]. 

Since many so-called 1/f noise are only ``1/f-like", meaning 
these spectra are not perfect power-law functions, it is very likely 
that the requirement in Eq.(\ref{eq24}) is too strong.
A reasonably broad distribution of correlation 
lengths might explain the data equally well. 

This point made here is very important to the understanding of 
the long-range correlation in DNA sequences. There is nothing
magic about the long-range correlation or 1/f spectra which could 
in principle be explained by the co-existence of many different length 
scales. What is essentially needed is a broad distribution of these 
different length scales (to be discussed more in comment \# 3 of  sec.4). 
Although this fact is well known in the 1/f-noise community, its 
relevance to the correlation structure in DNA sequence  should be
emphasized here.

\end{enumerate}

\subsection{Spectral analysis of DNA sequences}

Similar to the direct calculation of correlation function of DNA 
sequences, the application of spectral analysis to DNA is straightforward. 
As mentioned in the previous section, the advantage of spectral 
analysis is to reveal patterns hidden in a directly correlation function. 
But the early applications of spectral analysis were mainly 
focused on revealing  periodic signals
[McLachlan and Karn, 1983; Tavar\'{e} and Giddings, 1989,
and the references therein].
Only recently, attention turned to the functional shape of the power
spectrum at all frequency ranges [Li, 1992; Li and Kaneko, 1992a;
Voss,1992; Buldyrev, et. al. 1995]. Following two comments are
related to this topic:
\begin{enumerate}
\item
{\bf 1/f-like power spectra were observed in DNA sequences}: 1/f-like
power spectra were indeed observed in DNA sequences 
[Li and Kaneko, 1992a; Voss, 1992; Li, et. al., 1994]. These are
not white noise, indicating the existence of correlation. These are
not Lorentzian spectra either, indicating that there is a 
broad distribution for the correlation lengths in these
sequences. It was questioned in [Bor\u{s}tnik, et. al., 1993]
whether the ``apparent" $1/f^a$ spectra are actually Lorentzian spectra.
By the comment \#6 in the sec. 2.4, we see that superposition
of Lorentzian spectra can lead to a 1/f-like spectrum. 
The question is whether there is a single length scale 
(as in the case of Lorentzian spectrum) or multi-length scales 
(as in the case of 1/f spectra). 

\item
{\bf The quality of the 1/f spectra differs greatly among sequences}:
By the ``quality" of a 1/f spectrum, I mean a measure on how
close the observed spectrum is to a perfect $1/f^a$ ($a \approx 1$)
spectrum. As mentioned in comment \#1 in sec 2.3, different DNA sequences
do not necessarily exhibit the same power spectrum. The comment \#3 in
sec 2.3 says that the spectrum obtained by averaging many sequences 
(e.g., such as [Voss, 1992; Buldyrev, et. al. 1995]) may have a different
spectral form from that obtained in an individual sequence. 

The exponent $a$ in $1/f^a$ in [Voss, 1992] is very close to 1 (note
he subtracted the flat spectrum from the overall spectral shape), whereas
the $a$ in [Buldyrev, et. al. 1995] is much smaller: another indication
of the wide variation of the quality of 1/f spectra. The power spectra
of all sixteen budding yeast chromosomes are strikingly similar, all
1/f-like (W.Li, unpublished results and paper in preparation). 
The question still remains that to what extent the spectra of different
DNA sequences are similar or different to each other.

\end{enumerate}

\subsection{Other measures of the correlation structure}

\begin{itemize}

\item
{\bf DNA walk}:
In a controversial paper [Peng, et. al. 1992] (controversial
because some results could not be reproduced on the same data set
by two other groups [Prabhu and Claverie, 1992; 
Chatzidimitriou-Dreismann and Larhammar, 1992], a DNA sequence 
is first converted to a binary sequence (for example,
G or C is converted to 1, A and T converted to 0), then the binary
sequence is converted to a walk (1 for moving up, 0
for moving down). A random binary sequence leads to a random walk. 
The variance of this walk at certain distance $N$ is
related to the correlation function of the original binary sequence
[Peng, et. al. 1992; Karlin and Brendel, 1993]:
\begin{equation}
Var(N) =  N \Gamma_{binary}(d=0) + 2 \sum_{d=1}^{N-1} (N-d) \Gamma_{binary}(d)
\end{equation}
For random sequences, $\Gamma_{binary}(d)=0$ when $d >1$,
so only the first term is non-zero. It leads to
$Var(N) \sim N $. Any deviation from the linear relation indicates
a deviation from the random sequence.

One might be curious about whether the above relation can be generalized 
to cross correlation functions $\Gamma_{\alpha\beta}$ ($\alpha \ne \beta$). 
Indeed, using the identity
$\Gamma_{\alpha\beta}(-d) = \Gamma_{\beta\alpha}(d)$, we have:
\begin{equation}
Cov(N)_{\alpha\beta} = N \Gamma_{\alpha\beta}(d=0)
+ \sum_{d=1}^{N-1} (N-d)
[\Gamma_{\alpha\beta}(d) + \Gamma_{\beta\alpha}(d)]
\end{equation}
where $Cov_{\alpha\beta}$ should be the covariance of two such
converted walks, one for symbol $\alpha$ and another for symbol $\beta$.
When $\alpha \ne \beta$, $\Gamma_{\alpha\beta}(d=0) = - P_\alpha P_\beta$ 
is negative. Instead of linear increase, we have the case of linear decrease.

Converting a binary sequence to a walk is equivalent to carrying out
an integral or summation. Consequently, the statistics obtained from
the walk (such as the variance) are related to the integral (summation)
of the statistics from the original binary sequence (such as the
correlation function). The integral (summation) makes the statistics
from the walk smoother. Whether it is an advantage or disadvantage
depends on the purpose of the study. Actually, by using
a better estimator or by averaging many sequences, the direct
calculation of correlation function can also be very smooth.

\item
{\bf Graphical representation of DNA sequences}:
Similar to ``DNA walk", there have been many other proposals to map a 
DNA sequence to a graph, making visualization of the base density or 
correlation easier. Rather than reviewing these graphical representations, 
let me give a pointer to the original references: [Hamori and Ruskin, 1983; 
Hamori, 1985; Gate, 1986; Hamori, 1989; Hamori, et. al. 
1989; Jeffrey, 1990;  Berthelsen, et. al., 1992; Pickover, 1992; 
Wu, et. al. 1993; Zhang and Zhang, 1994].

Another type of graphical representation concerns the sequence-dependent
bending/curving [Calladine and Drew, 1990].  These graphical 
representations of a DNA sequence require some biochemical modeling 
of the spatial bending of the double helix of the DNA molecule 
(e.g. [Tung and Harvey, 1986; Shpigelman et. al. 1993; Tung and Carter, 1994]). 
A review of this topic is outside the scope of this paper.

\item
{\bf Wavelet analysis}:
The wavelet transformation [Daubechies, 1988;1992] 
is a new type of transformation where localized ``wavelets" 
replace the sine/cosine functions in the Fourier transform as
the basis. This new method is ideal for studying heterogeneity
in DNA sequences, a topic to be discussed in the next subsection.
For references on wavelet transformation applied to DNA sequences,
see [Zhang, 1995; Arneodo et. al., 1995; 1996]. 

\end{itemize}

\subsection{Study of heterogeneity}

As pointed out in comment \#1 in sec 2.1, statistics are sampled
along DNA sequences. If there is heterogeneity along the sequence, 
it is averaged out during the sampling process. This heterogeneity is 
intrinsically related to the slow-decay of correlation function: 
in order to have a significant correlation at longer distances, 
the correlated units should be larger than a few bases.  These 
larger units, relatively homogeneous sub-sequences, can be called 
``domains". 

Heterogeneity of base density in DNA sequences was observed 
in digested DNA fragments (e.g., [Sueoka, 1959; Filipski, et. al. 1973; 
Thiery, et. al. 1976;  Macaya, et. al.  1976;  Cuny, et. al. 1981]).  There
were limitations in this experimental determination of the heterogeneity. 
First, only the C+G density of one DNA fragment is determined. The
spatial variation of C+G density within the fragment cannot be detected.
Second,  though a C+G-rich fragment is usually followed by a
C+G-poor fragment, it is not known exactly how these fragments
are assembled to the original long DNA sequences, so the correlation
among these fragments is  not well known.

These experimental studies inspired a few theoretical studies
[Elton, 1974; Churchill, 1989; Kozhukhin and Pevzner, 1991;
Fickett, et. al., 1992]. The modeling of sequence 
heterogeneity will be discussed in the next section. 
Here, let me address the issue on how to {\sl characterize} the 
heterogeneity in a DNA sequence. 

Consider the simplest situation of heterogeneity in a DNA sequence:
the left half of the sequence is highly C+G rich and the right half
is C+G poor. A characterization of the heterogeneity includes the 
information on the boundary between the two homogeneous halves, 
the difference of C+G density (or in general base densities) between 
them, the size of each domain, etc. Some potential problems can be seen 
immediately: what if the C+G density on the left half 
is only slightly different from that in the right half --
do we still consider the sequence to consist of two domains?
What if the left half can be decomposed to sub-domains itself? 
The problem is that we may not always have a clear-cut domain structure.

Methods that identify and partition the whole sequence into
homogeneous domains are {\sl segmentation algorithms}, a term used
in image processing. The basic idea in a segmentation algorithm is
to use a measure of the degree of fluctuation (such as variance),
find a partition point which minimizes the fluctuation in either
partition as compared with the degree of fluctuation in the original
unpartitioned sequence. 

One conceptually simple segmentation 
algorithm was proposed and applied to DNA sequences
[Bernaola-Galv\'{a}n, et. al.,  1996]. 
In this segmentation algorithm, the single-base entropy, 
$H_1 \equiv - \sum_{\alpha =(A,C,G,T)} P_\alpha \log_2 P_\alpha $, 
is used to measure the fluctuation ($H_1$ measures the randomness 
at the base level). The weighted sum of the two entropies 
from both partitions is compared with the overall entropy
from the original sequence. A partition point 
is determined when the difference between the two (called 
``Jensen-Shannon divergence" in [Lin, 1991]) is maximized.

An extension to the above approach is to use high-order entropies
(e.g. $H_2 \equiv -\sum_{\alpha\beta} P_{\alpha\beta} \log_2 P_{\alpha\beta}$
for 2-base entropy, $H_N$ for $N$-base entropy,
and the source entropy $h \equiv \lim_{N \rightarrow \infty} H_N/N$
[Shannon, 1951]). 
Typically, high-order entropies are difficult to calculate 
due to lack of sample size. One approach to extrapolate to
the infinite-size limit is to use the regression analysis
[Konopka, 1994]. Another alternative is to use some measure 
of compressibility [Ziv and Lempel, 1977] which is intrinsically related 
to the source entropy. Both approaches are subject to 
finite-sample-size effect and may not give a reliable estimation 
of the source entropy. In the case of DNA sequences it has
indeed been suggested [Konopka, 1994] that extrapolated
source  entropy carries a high sampling error and thereby
is not a useful criterion to discriminate between introns
and exons.

There is an important parameter to be decided in a segmentation algorithm,
which is the threshold of the divergence. In essence, the question is
how different the two domains should be before we distinguish the two. 
A low threshold value makes it easier to partition, even for random sequences. 
A high threshold value leads to no partition at all. Changing the 
threshold value, we may change the size distribution of the partitioned 
domains. In [Bernaola-Galv\'{a}n, et. al., 1996], the
size distributions of different DNA sequences are compared and
quantitative differences were observed. For future studies, such 
comparison might be carried out at each of the possible threshold values. 
A family of the size distributions is then compared with another
family of distributions.

A crucial question concerning the heterogeneity of a DNA sequence
can be answered by these segmentation algorithms: whether there
are domains within a domain. For simple heterogeneity, there are
no domains within a homogeneous domain. For complex heterogeneity,
the concept of homogeneity is only relative: a domain which is
homogeneous under one threshold value can be heterogeneous under another
slightly lower threshold value. This phenomenon was indeed observed
in [Bernaola-Galv\'{a}n, et. al.,  1996] for
one sequence, and not in another sequence. The claim that ``the mosaic
character of DNA consisting of domains of different composition
can fully account for apparent long-range correlations in DNA" 
[Karlin and Brendel, 1993] underestimates the true complexity of the
heterogeneity problem in DNA sequences.

\section{Modeling DNA sequences}

Characterizing the correlation structure of DNA sequences does not
involve any modeling of the sequence. The observation of the
slower-than-exponential decay of correlation function and the 1/f-like
power spectra in DNA sequences does not require any assumption about
the sequence. We notice a comment in [Karlin and Brendel, 1993]: 
``Recent papers proffer the asymptotically weakly independent stationary 
process as a model to describe apparent long-range dependencies inherent to
many DNA sequences. However, the assumption of stochastic stationarity 
is problematic $\dots$". This comment confuses the two:  by calculating
the correlation function or the power spectrum, we only extract
information from the DNA sequence and summarize this information in a compact
form; we do not automatically assume the DNA sequences to be homogeneous
(stationary).  Four different modeling strategies of DNA sequences
are reviewed below. 

\subsection{One-step Markov chains were known to be a poor model 
for DNA sequences}

One of the earliest attempts to model DNA sequences was to use
the 1-step Markov chain  [Gatlin, 1966, 1972; Elton, 1974]). It was then
realized that 1-step Markov chains are not good model for DNA 
sequences. One reason is that in coding regions, the position
within a codon is important, thus the Markov transition probabilities
depend on this position [Borodovskii, et. al. 1986a, 1986b; Tavar\'{e}
and Song, 1989]. This problem is relatively easier to fix: one might use
Markov models with cyclic, position-dependent, transition probabilities. 
Another reason is due to heterogeneity:  on the global scale the 
Markov transition probabilities depend on which homogeneous domain the
position falls in [Borodovskii, et. al. 1986a, 1986b]. This problem can
not be solved within the framework of a Markov model.

The correlation structure of DNA sequences gives another confirmation
that 1-step Markov chains are not good models for DNA sequences. 
The correlation function for a 1-step Markov chain decays exponentially. 
This can be easily proved: the joint probability $P_{\alpha\beta}(d)$ 
can be obtained by calculating the $d$'th power of the Markov transition 
matrix, which is dominated by the $d$'th power of the largest eigenvalue
of the matrix. And this leads to an exponential decay for
$P_{\alpha\beta}(d)$ (see, e.g., [Feller, 1968]).
Since we know from sec 2.3 that the correlation function in DNA
sequences does not usually decay exponentially, then a 1-step Markov
chain does not characterize the observed correlation functions. 

\subsection{High-order Markov chains are penalized by having too many
free parameters}

High-order Markov chains certainly  have
more degrees of freedom to characterize a wider variety of correlation 
structures than the 1-step Markov chains. In particular, with multiple
eigenvalues for the transition matrix (each related to a different decay
rate of the exponential function), it is not impossible (though
practically unlikely)  to have such a
distribution for these eigenvalues that the resulting mixture of 
exponential functions lead to a 1/f-like power spectrum [Li, 1987]. 

Nevertheless, there is a major drawback in using high-order Markov
chains: the number of free parameters increases with the square of
the order of the Markov chain. In the Bayesian framework of model
selection, models with more free parameters are penalized (see, e.g. 
[Sivia, 1996]), in the spirit of the ``Ockham principle". If one has
to use a model with many parameters, it must fit the data comparatively 
much better to compensate for this penalty.  In [Raffery and Tavar\'{e}, 1994],
a certain assumption is made concerning relationship among the parameters
in the model, thus reducing the number of independent parameters. But
this assumption might not be justified. 

The fundamental reason which casts doubt on 1-step Markov model
also negates the high-order Markov chains: as long as the order of 
the Markov chain is finite, it will not characterize the heterogeneity 
at the length scale larger than its order.

\subsection{Hidden Markov chains are ideal model for simple heterogeneity}

The obvious heterogeneity in DNA sequences motivated the introduction
of other mechanisms that possibly describe the phenomenon. 
One powerful model is the hidden Markov chain or hidden Markov model
[Baum and Petrie, 1966; Rabiner, 1989]. There are two layers of variables
in a hidden Markov model.  When it is applied to DNA sequences 
[Churchill, 1989; 1992], these two layers of variables
are:  (1) the observed variable, which is the four 
nucleotides along a sequence $\{ O(i) \}$ ($i=1, \dots N$); 
and (2) the state variable $\{ S(i) \}$ (unobserved), which is
related to a description of the domain. 

The state variable sequence can be produced by a Markov chain. 
Thus, a hidden Markov model is characterized by two sets of transition
probabilities: one is the state transition probability 
$Prob(S(i) \rightarrow S(i+1))$ (suppose the Markov chain is 1-step); 
another is the mapping probability from the state variable to 
the observable $Prob(S(i) \rightarrow O(i))$.

The state variable can be discrete or continuous. For an example
of the discrete state variable, consider three possible values
for $\{ S(i) \}$: $H, M, L$, for high, moderate, and low C+G density. 
The mapping probabilities are different for different state variables.
For example, when $S(i)=H$, then 
$Prob(H \rightarrow G)$ and $Prob(H \rightarrow C)$ 
are higher than $Prob(H \rightarrow A)$ and $Prob(H \rightarrow T)$.

For an example of the continuous state variable, consider $S(i)$ 
being the C+G density at the site $i$, which can take any real number
between 0 and 1 (in practice, the lower and upper limits
can be 1/3 and 2/3 [Fickett, et. al., 1992]). With a given
value of $S(i)$, say, $s$, the mapping probability can be chosen as
$Prob(S(i) \rightarrow C) = Prob(S(i) \rightarrow G) = s/2$, 
$Prob(S(i) \rightarrow A) = Prob(S(i) \rightarrow T) = (1-s)/2$. 
The continuous state variable
version of the hidden Markov chain was called ``walking Markov chain"
in [Fickett, et. al., 1992]. Comparing the continuous state variable
version with the discrete version, the former does not necessarily
lead to a sharp boundary between domains.

Similar to the case of high-order Markov chain, there is also an
issue of estimating many parameters in the hidden Markov chains,
thus the fitting of the data must be very good in order to
compensate the penalty for using too many parameters. But overall, 
hidden Markov chains are a much better choice for characterizing
heterogeneity in DNA sequences than the high-order Markov chains.

\subsection{Rewriting systems and complex heterogeneity}

Hidden Markov chain represents a major improvement over the
regular Markov chain in characterizing DNA sequences because
it is able to describe the heterogeneity well. But as discussed
in sec 2.7, when the heterogeneity becomes complex, a seemingly
homogeneous region can be heterogeneous when the criterion for
being heterogeneous is relaxed. The same phenomenon may reappear
at a subdomain level. In other words, there can be subdomains within a
domain, sub-sub-domains within a sub-domain, etc.

Rewriting systems are able to generate such hierarchical organization
of domains and self-similarity. Being rediscovered and  renamed a few
times,  such as the context-free language [Chomsky, 1956] (if
the distinction between non-terminal symbols and terminal symbols is
removed), L-systems [Lindenmayer, 1968; Rozenberg and Salomaa, 1980;
Prusinkiewicz and Lindenmayer, 1990], 
development systems [W\c{e}grzyn, et. al. 1990], substitutional sequences
[Cheng and Savit, 1990], a rewriting system updates a sequence by replacing
a symbol (or a string) by another string of symbols.

For example, the famous Fibonacci sequence is generated
by repeated use of the following rewriting rule" (replacing a symbol 
at $t$ with a string of symbols at $t+1$):
\begin{equation}
0_t \rightarrow 1_{t+1} \hspace{0.5in} 1_t \rightarrow (10)_{t+1}
\end{equation}
The expansion-modification system [Li, 1989; 1991]  is a 
probablistic rewriting system:
\begin{equation}
0_t \stackrel{1-p}{\rightarrow} (00)_{t+1}
\hspace{0.3in}
0_t \stackrel{p}{\rightarrow} (1)_{t+1}
\hspace{0.3in}
1_t \stackrel{1-p}{\rightarrow} (11)_{t+1}
\hspace{0.3in}
1_t \stackrel{p}{\rightarrow} (0)_{t+1}
\end{equation}
with the probability attached to each rule.
One might easily generalize this model to a model
of duplication and point mutation in DNA sequences, such as
\begin{equation}
T_t \rightarrow A_{t+1}, 
\hspace{0.4in}  (ACC)_t \rightarrow (ACCACC)_{t+1}
\cdots, etc.
\end{equation}
If there is conflict among different rewriting rules, we might
choose each one with certain probability.

Rewriting systems operate quite differently from Markov chain
and its variants. Markov chain moves along a sequence and
generate new symbols sequentially. Rewriting systems not only
update symbols {\sl parallelly}, but also 
{\sl repeatedly}. These two features
give rewriting systems certain advantages over Markov chains
in modeling long DNA sequences which might result from
a long evolutionary process involving repeated duplications.

There are also several parameters to be determined in a rewriting 
system. For example the initial sequences, the number of 
times the rule is applied, and the probability of applying certain
rules when there is a conflict. It would be interesting  to
use a more rigorous model selection procedure to choose
the type of rewriting system that might characterize DNA
sequences well, and to compare the rewriting system models with
the hidden Markov models.

\section{Some biological issues}

Is the study of correlation structure of DNA sequences useful
for working biologists? Since most of the current DNA sequence 
analysis is based on the knowledge of local signals (e.g. consensus 
pattern of regulatory regions), the global view of the DNA sequences 
takes a back seat. Let me use an analogy of the natural language 
texts. Understanding local signals is similar to the construction 
of a dictionary of words. The interaction among different regulatory 
regions is perhaps analogous to syntax of sentences. The evolutionary
history of the DNA sequences is analogous to the writing of a text 
(with repeated redrafting).  And the resulting genome organization 
is analogous to a style of the text. If a dictionary has already been 
constructed, the next goal should be an understanding of the whole text.

Here are a list of biological issues related to the study of correlation
structure of DNA sequences:

\begin{enumerate}

\item
{\bf Concentration of genes and the C+G density}:

It was shown that the concentration of genes is correlated with
the C+G density [Bernardi, et. al. 1985; Bernardi, 1989; 1995; 
Zoubak, et. al., 1996; Clay, et. al. 1996].  The concentration of genes
in the C+G-richest region of human genome can be 5-10 [Bernardi, 1989]
or  15 times [Zoubak, et. al., 1996] higher than those of the
C+G-poorest regions.  If the connection between C+G density and 
the concentration of genes is proven general, spatial distribution
of C+G density can be used to give an indication of the locations of genes.

The base-base correlation structure reviewed here can be extended 
easily to the density-density correlation. One can partition a sequence 
into  {\sl non-overlapping} windows, and calculate the C+G density 
in each window.  Similarly, the spatial distribution of genes can
be characterized by the binary sequence which assigns a 1 to a base
in the gene and 0 to a base in the intergenic region. The cross-correlation
between the density sequence and the binary sequence can answer
quantitatively the question of whether the spatial distribution of genes
is related to the spatial variation of C+G density.

\item
{\bf Besides the period-3 structure in coding region, are the
correlation structures in coding and non-coding regions different?}:

Due to the codon structure, the correlation function for coding sequences
exhibits period-3 structure with a high-low-low pattern [Fickett, 1982].
Eukaryotic intron sequences do not exhibit this period-3 pattern, but 
occasionally, they exhibit a period-2 pattern [Konopka and Smythers, 1987; 
Konopka, et. al. 1987; Arqu\`{e}s and Michel, 1987; Li, 1992] (unlike nuclear
DNA sequences, the non-coding regions in viral or mitochondrial 
sequences may still have period-3 patterns [Arqu\`{e}s and Michel, 1987;
Konopka, 1994]).

Besides this well-known pattern, how does correlation function decay with 
distances in either coding and non-coding sequences, and how do 
the correlation structures differ in the two types of sequences? 
Despite several studies, this remains an open question. 
Intuitively, because of the specific constraints imposed by the 
function of coding region, i.e., the 3-nucleotide to 1-amino-acid 
translation as well as the ability for the translated amino acid sequence 
to fold successfully, evolutionary tinkering is mostly prohibited in the coding 
region. On the other hand, non-coding regions (both intron and the intergenic 
regions) are open to changes, because these changes typically do not lead
to fatal damage to the cell and the organism.

One of the most important changes in DNA sequences is 
duplication [Bridges, 1919; chapter 17 of Morgan, et. al.,  1925],
either the oligonucleotide duplication or gene duplication. The gene
duplication in particular, is considered to be crucial for the 
generation of complexity and the acquisition of new functions
in high organisms [Ohno, 1970; Market, et. al., 1975; 
MacIntype, 1976; Doolittle, 1981].  To quote from 
[Doolittle, 1981]: ``\dots it is simpler to duplicate and modify 
proteins genetically than it is to assemble appropriate
amino acid combinations de novo from random beginning".

If the changes in non-coding sequences are mainly duplications, the non-coding
regions should be less-random, or more-redundant.  This would predict a 
longer-ranged correlation in non-coding sequences. Indeed,  [Li, 1992] 
showed that this is true for a few human DNA sequences. But [Voss, 1993; 1994] 
did not observe significant difference in power spectra between coding and 
non-coding sequences. More recently, another study shows that after the 
period-3 structure is subtracted, the mutual information functions for 
coding and non-coding regions are very similar to each other, both 
decay as an approximate power-law function in 1-1000 bases range 
(I. Gro\ss e, private communication, 1996). This study paid particular 
attention to correcting the bias in the estimation due to finite 
sequence length to make sure the small correlation at long distances 
is estimated accurately.

So which study is correct? In light of the fact that the currently available 
DNA sequences are biased towards coding sequences and the flanking 
sequences, I would like to caution that once the intergenic sequences are
represented in the sample with a larger proportion, the conclusion might
be modified accordingly. Intergenic sequences may expose to a lesser degree of 
constraints as compared to the intron sequences (with perhaps the 
exception of structural constraints), thus exhibiting different statistical
features. Indeed, intergenic sequences are analyzed separately from
the intron sequence in, for example, [Guig\'{o} and Fickett, 1995]).  

Human intergenic sequences, in particular, have more ``room" to 
accumulate changes due to their sheer size.  With the combination of 
the two (less constraint and larger size), one can imagine that duplication, 
insertion and point mutation can play a major role in shaping the 
intergenic sequences (this speculation is already supported by a study 
of intergenic sequences in maize genome [SanMiguel, et. al. 1996]). 
Thus, human intergenic sequences may more easily exhibit long-range
correlation than other types of sequences.

\item
{\bf Broad distribution of domain sizes as the origin 
of power-law decay of correlation function}:

There have been many papers discussing the ``biological origin"
of long-range correlations. But what was mostly discussed was 
the known fact that the base density may be different 
at different regions, i.e., the heterogeneity. What is non-trivial
about the correlation structure in many DNA sequences should be
discussed along the line of ``complex heterogeneity versus simple 
heterogeneity", ``$1/f$ spectra versus $1/f^2$ spectra", 
``broad versus narrow distribution of domain sizes".

For example, for the DNA sequence of a complete chromosome, the sequence
can be partitioned into coding and non-coding domains. Assuming 
coding and non-coding domains are distinct in base density,
the size distribution of coding as well as non-coding domains determines
the correlation structure of the whole sequence. A study along
this line was presented in [Herzel and Gro\ss e, 1997] for budding
yeast chromosomes, and a broad distribution for coding domains 
(actually it is the distribution of the ``open reading frames", 
i.e., the potential or putative coding regions). 
Note that this discussion does not apply to
the correlation structure in intergenic sequences, and the assumption
that a clear distinction between the base densities in coding and
non-coding domains may not always hold.

One can attribute the broad distribution to either internal or 
external factors. For external factors, it was shown that the size 
distribution of insertions and deletions of pseudogenes is approximately 
power-law functions [Gu and Li, 1995]. If these insertions are distinct
base-density-wise from the hosting DNA sequence, the power-law 
distribution of the sizes is enough to lead to a complex 
heterogeneity.  The power-law distribution of insertion sizes was used 
in  [Buldyrev, et. al. 1993] by the assumption that the insertion is 
accomplished by a loop formation, and the size distribution of the 
loop length in a long polymer should be a power-law function [des 
Cloizeaux, 1980]. But it has not been examined whether this argument is 
correct.

Duplication is a best example of the internal factors for
broad distributions. The intuition can be again derived from the 
expansion-modification system [Li, 1989; 1991].  Since a mutation
followed by the repeated duplication of that mutated base generates 
a distinct domain whose size is proportional to the number of 
duplications, in order to have a broad distribution of sizes, one 
simply needs different duplication events to start at different times. 
Then the broad distribution is just a historical profile of
these duplication events. For a more realistic modeling of the
(gene) duplication events, see [Ohta, 1987a; 1987b; 1988a; 1988b; 1989; 1990]

Using oligonucleotide repeats as an explanation of the 1/f-like spectra
in DNA sequence [Li and Kaneko, 1992b] does not mean that a  single
simple repeat can explain the correlation structure in DNA sequences: 
it will not, because such a repeat manifests itself as a peak in the power 
spectrum instead of a broad band 1/f spectrum.  Rather, what we suggested 
in [Li and Kaneko, 1992b] was a series of duplications each occurring 
at a different historical period, and repeating a variety of times. 

Finally, it is also possible that both external and internal factors 
contribute to the broad distribution of domain sizes.

\item
{\bf Correlation between other units}:

There can be many variations to the average correlation between two 
bases calculated along a sequence. One example is the statistical 
correlation between two specific sites. The statistic in this case 
is no longer sampled along a sequence, but from a set of aligned 
sequences at these two specific sites.  For example, the mutual 
information between two specific sites in HIV proteins was calculated 
to detect the co-varying mutation spots [Korber, et. al. 1993]. 
In this example, a set of similar sequences is made possible by
the high variability of the HIV. 

If there are only a few copies of a biologically meaningful unit
on a DNA sequence, it is difficult to obtain a statistic.  Take the 
replication origins on budding yeast chromosomes for example 
[Newlon, et. al. 1993]. The replication origins could be determined 
through a sequence analysis by searching the 10- or 11-base concensus 
pattern [Palzkill, et. al., 1986] (this might be called putative
replication origin because of the possibility of false positives and
the possibility of inactivation of the origin). One yeast chromosome 
may have only a limited number of replication origins (around the order 
of 10). Thus, a proposition concerning these origins may be simply
studied by an exhaustive listing without using any statistics. For example,
in order to test the proposition that replication origins are always
located inside a non-coding region [Murakami, et. al. 1995], we might just
count how many replication origins are indeed located in the non-coding region.

\end{enumerate}

\section*{Conclusion}

Determining correlation structure of DNA sequences on a more global scale
reveals that the picture of simple heterogeneity is not enough to 
explain features of many sequences. A key to the understanding of the
complexity of correlation structure in DNA sequences is the broad
distribution of domain sizes. More quantitative measurement is
necessary to characterize this feature more accurately. With even longer
stretches of continuous DNA sequences to be available in the future, it is 
conceivable that correlation structure at even larger scales can be 
studied.  Blurring the details at base level might be necessary in order 
to detect any significant correlation at these larger scales. Our grand 
goal is to eventually learn the ``genome organization" principles, and
explain this organization using our knowledge about evolution. On
the latter aspect, this study may look similar to the study of
molecular evolution and population biology (see, e.g.,
[Kimura, 1983; Gillespie, 1991]), but with more emphasis on
spatial correlation in DNA sequences.  An online bibliography on the 
topic of of correlation structure 
of DNA sequences can be found at: 
{\sl http://linkage.rockefeller.edu/wli/dna\_corr}.

\section*{Acknowledgements}

This review would be impossible without extensive discussions with Ivo 
Gro\ss e, Andrzej Konopka, J\'{o}se Oliver, Oliver Clay, and Michael 
Teitelman, and I would like to thank them all. I also thank Tara Matise 
to proofreading the draft. A partial support from the Aspen Center 
for Physics for attending the ``Identifying Features in Biological 
Sequences (1996)" is acknowledged. The author is supported by the 
grant HG00008 (to J. Ott) from the National Human Genome Research 
Institute of NIH.

\vspace{0.25in}


\newpage

\begin{figure}
\begin{center}
\epsfig{file=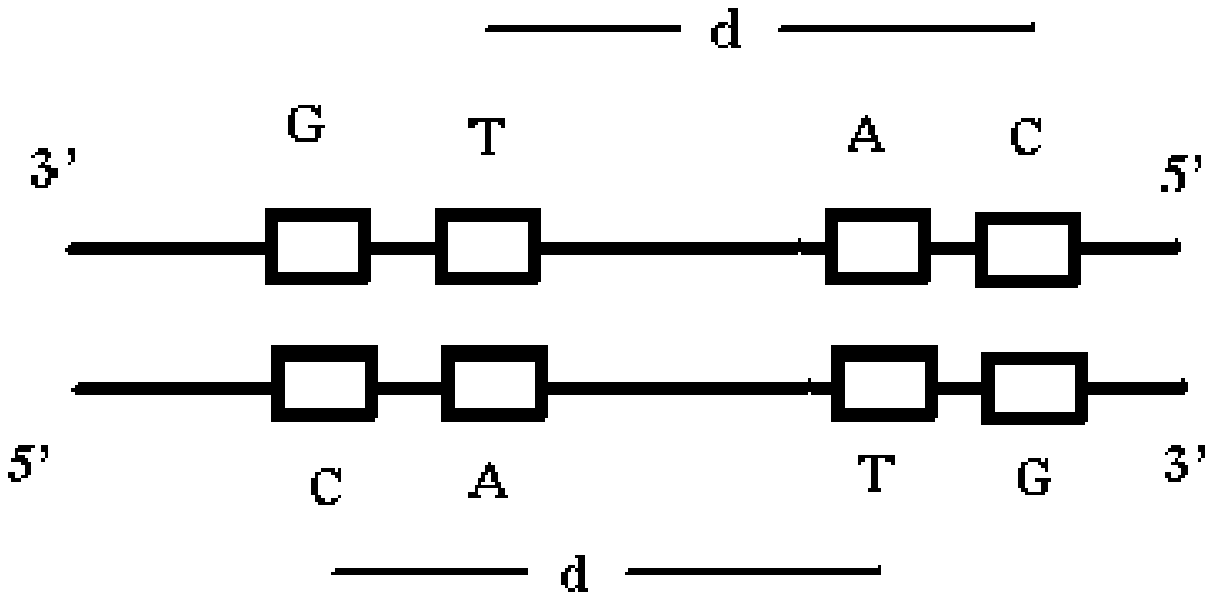}
\end{center}
\caption{
Illustration of the strand complementary and strand symmetry.
If the 5'-C-T-3' correlation on the lower strand is the same
with the 5'-C-T-3' correlation on the upper strand, by
complementarity, 5'-C-T-3'  and 5'-A-G-3' both on the lower
strand are the same.
}
\label{fig1}
\end{figure}

\begin{figure}
\begin{center}
\epsfig{file=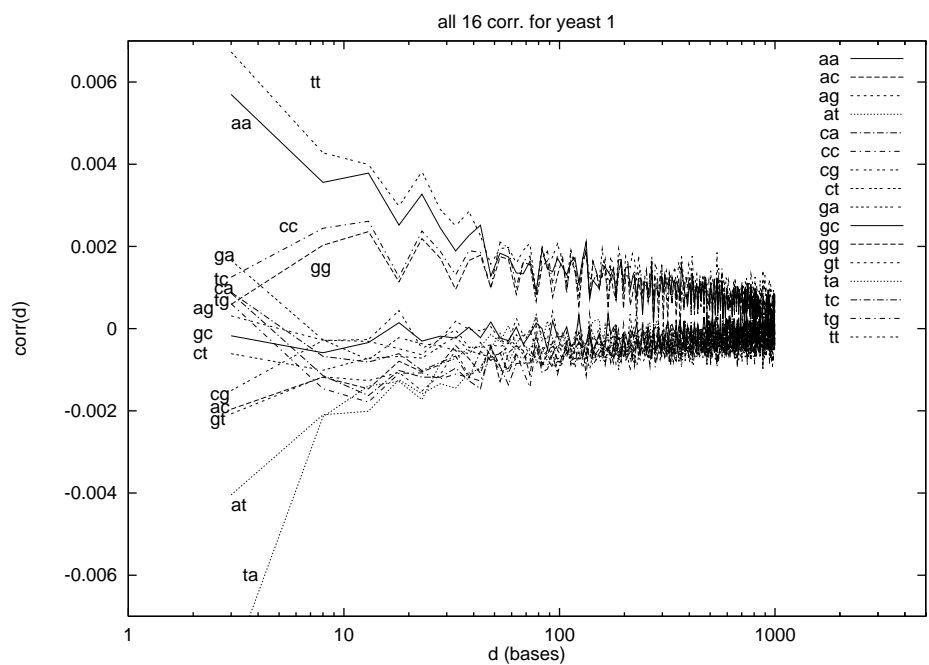}
\end{center}
\caption{
All 16 correlation functions for budding yeast chromosome 1 sequence.
Correlations at 5 neighboring distances (e.g, d=1,2,3,4,5) are averaged
to smooth the curve. The Bayesian estimator of the correlation
function  (Eq.(\ref{eq15})) is used. 
}
\label{fig2}
\end{figure}

\end{document}